\documentclass[final,3p,twocolumn,authoryear]{elsarticle}
\usepackage{graphicx}
\usepackage{amssymb}
\usepackage{amsmath}

\begin{document}
\begin{frontmatter}


\title{Lagrangian study of transport and mixing in a mesoscale eddy street}


\author{S.V. Prants}
\ead{prants@poi.dvo.ru}
\ead[url]{www.dynalab.poi.dvo.ru}

\author{M.V. Budyansky}
\author{V.I. Ponomarev}
\author{M.Yu. Uleysky}

\address{Pacific Oceanological Institute of the Russian Academy
of Sciences, \\ 43 Baltiiskaya st., 690041 Vladivostok, Russia}

\begin{abstract}
We use dynamical systems approach and Lagrangian tools to study surface
transport and mixing of water masses in a selected coastal region of the
Japan Sea with moving mesoscale eddies associated with the Primorskoye Current.
Lagrangian trajectories are computed for a large number of particles in an
interpolated velocity field generated by a numerical regional multi-layer
eddy-resolving circulation model. We compute finite-time Lyapunov exponents
for a comparatively long period of time by the method developed and plot the Lyapunov
synoptic map quantifying surface transport and mixing in that region.
This map uncovers the striking flow structures along the coast with a
mesoscale eddy street and repelling material lines. We propose new Lagrangian
diagnostic tools --- the time of exit of particles off a selected box, the
number of changes of the sign of zonal and meridional velocities --- to
study transport and mixing by a pair of strongly interacting eddies often
visible at sea-surface temperature satellite images in that region.
We develop a technique to track evolution of clusters of particles,
streaklines and material lines. The Lagrangian
tools used allow us to reveal mesoscale eddies and their structure, to track
different phases of the coastal flow, to find inhomogeneous character
of transport and mixing on mesoscales and submesoscales and to quantify
mixing by the values of exit times and the number of times particles
wind around the eddy's center.
\end{abstract}

\begin{keyword}
Lagrangian transport \sep mesoscale eddy dynamics \sep Lyapunov exponents


\end{keyword}
\end{frontmatter}

\section{Introduction}

Large-scale coherent structures, such as major currents and eddies, are key
ingredients organizing ocean flows. They are easily seen on surface
temperature and sea-surface height satellite data. In numerical models,
contours of potential vorticity can be used to reveal coherent structures.
However, those and other Eulerian means give snapshots at fixed times and
contain little information about transport and mixing of water masses which
are inherent Lagrangian notions. Lagrangian approach to study horizontal
transport and mixing in the ocean is rapidly becoming an effective method
to demonstrate inhomogeneous character of that mixing and existence of
coherent structures
\citep{Haller,Wiggins,KP06,H00,HY00,SL05,IS02,MS06,JW02,Samelson,
Artale,Koshel08,KVK08}.
There are interesting papers on that approach studying transport and mixing
in different basins in the World Ocean with numerically generated velocity
fields \citep{BO08,OF04,MH08,OR06,GI07,Miller} and velocity fields derived
from satellite altimeter measurements \citep{OI09,WA06,AR02,LO07}
or using high-frequency radars \citep{LC05,LS06,GF09}.
This approach does not aim at studying individual trajectories of fluid
particles but at searching for and identifying spatial structures organizing
the whole flow and known in theory of dynamical systems as invariant manifolds.

Motion of a fluid particle is the trajectory of a dynamical system with  given
initial conditions governed by the velocity field computed either by solving the
corresponding master equations or as the output of a numerical ocean model or
derived from a measurement. The phase space of that dynamical system is the
real space where many important phase-space objects known in the theory of
dynamical systems, such as stationary points, Kolmogorov-Arnold-Moser tori,
stable and unstable manifolds, periodic and chaotic orbits, etc.,
can be found and studied.

In this work, we use velocity data from the Japan Sea circulation model
\citep{JSmodel} to study and characterize surface transport and mixing
in the region comprising the Primorskoye (Liman) Current which is known by its
rich mesoscale activity. The instabilities of this current in the warm period
generate mesoscale coastal eddies \citep{Ponomarev} which propagate
downstream until the entrance to the Peter the Great Bay (the latitude of
Vladivostok, Russia). These eddies are generated in the model as a
kind of a mesoscale eddy street. The mesoscale and submesoscale dynamics
over the shelf and steep continental slope includes jet currents, streamers
and eddies being controlled by synoptic scale wind forcing and sea
baroclinicity. According to satellite data, the anticyclonic mesoscale
eddies of relatively small scale of the order $(10 \div 50)$~km
have been observed in the northwestern marginal area directly over the steep
continental slope. Near the sea surface they often form
pairs of strongly interacting eddies with spiral patterns visible
at satellite images. The anticyclonic mesoscale eddies of larger scale
have been clearly seen in the southern marginal area of the slope and shelf of
the Peter the Great Bay. The mesoscale dynamics over the continental slope
could be associated with the coastal Kelvin waves propagating downstream
with the Primorskoye Current to the southwest and catched by the wide shelf
of the Peter the Great Bay.

Many practical transport problems focus on tracking the evolution of small
fluid patches (clusters of particles) near the coast that becomes important
in the case of oil and other pollutant spills or harmful algal blooms.
It is of crucial interest to know the fate of the particles inside the patch.
As to basic research problems in physical oceanography, it is important to
know typical transport pathways along which coastal waters move to the open
sea and the open-sea waters move to the coast.

The aim of this paper is to study surface transport and mixing along
the mesoscale eddy street by computing Lagrangian trajectories for a large
number of particles advected by the Japan Sea circulation model known as
the MHI model (Marine Hydrophysical Institute, Sevastopol, Ukraine)
\citep{JSmodel}. Our paper is organized as follows. In section 2, we
review briefly the dynamical systems method to study transport and mixing
in the ocean. Section 3 introduces the Japan Sea circulation model.
Section 4 contains our main results. (i) We compute  finite-time Lyapunov
exponents (FTLE),
the time of exit of a large number of particles off a selected sea region and
streaklines. It enables us to reveal mesoscale eddies, hyperbolic and non-hyperbolic
regions in the sea and to quantify mixing by values of the FTLE.
(ii) We track the evolution of material lines, crossing the eddies,
and demonstrate inhomogeneous character of transport and mixing on mesoscales and
submesoscales that can be quantified by values of exit
times and the number of times particles change their zonal velocity while
moving in the preselected box. (iii) We show that the evolution of neighbour
patches of particles, chosen on a ridge of largest values of the FTLE field
and nearby, illustrates strongly different transport paths of particles
in hyperbolic sea regions. We summarize our results in
section~5.  In Appendixes A and B, we illustrate the complex pattern of chaotic
mixing and transport by an idealized example of the vortex flow
and describe the method of computing the FTLEs, respectively.

\section{Dynamical systems method to study transport and mixing in the ocean}

In Lagrangian approach, a fluid particle is advected by the two-dimensional
Eulerian velocity field
\begin{equation}
\frac{d x}{d t}= u(x,y,t),\quad \frac{d y}{d t}= v(x,y,t),
\label{adveq}
\end{equation}
where $(x,y)$ is the location of the particle, $u$ and $v$ are the zonal
and meridional components of its velocity at the location $(x,y)$.
The motion is considered on a plane because of a comparatively small
size of the region we will analyze in this paper. Even if the velocity
field is fully deterministic, the Lagrangian trajectories may be very
complicated and practically unpredictable. It means that a distance
between two initially nearby particles grows exponentially in time
\begin{equation}
\| \delta {\mathbf r}(t) \| = \| \delta {\mathbf r}(0) \|\, e^{\lambda t},
\label{Lyap}
\end{equation}
where $\lambda$ is a positive number, known as the Lyapunov exponent,
which characterizes asymptotically (at $t\to \infty$) the average rate
of the particle dispersion, and $\|\cdot\|$ is a norm of the vector
$\mathbf{r}=(x,y)$. It immediately follows from (\ref{Lyap}) that we are unable
to forecast the fate of the particles beyond the so-called predictability
horizon
\begin{equation}
T_p\simeq\frac{1}{\lambda}\ln\frac{\|\Delta \|}{\|\Delta (0)\|},
\label{horizon}
\end{equation}
where $\|\Delta  \|$ is the confidence interval of the particle location
and $\|\Delta (0)\|$ is a practically inevitable inaccuracy in
specifying the initial location. The deterministic dynamical system
(\ref{adveq}) with a positive maximal Lyapunov exponent for almost all
vectors $\delta \mathbf{r} (0)$ (in the sense of nonzero measure) is called
chaotic. It should be stressed that the dependence of the predictability
horizon $T_p$ on the lack of our knowledge of exact location is logarithmic,
i.~e., it is much weaker than on the measure of dynamical instability
quantified by $\lambda$. Simply speaking, with any reasonable degree of
accuracy on specifying initial conditions there is a time interval beyond
which the forecast is impossible, and that time may be rather small for
chaotic systems.

In the last two decades, dynamical systems methods have been applied to study
transport and mixing processes in the ocean
\citep{Haller,Wiggins,KP06,H00,HY00,SL05,IS02,MS06,JW02,Miller,Samelson,Artale}.
Since the phase plane of the two-dimensional dynamical system
(\ref{adveq}) is the physical space for fluid particles, many abstract
mathematical objects from dynamical systems theory are material surfaces,
points and curves in fluid flows. Stagnation point in a steady
flow is the fluid particle with zero velocity. Besides ``trivial'' elliptic
stagnation points, the motion around which is stable, there are hyperbolic
(saddle) stagnation points which organize fluid motion in their
neighbourhood in a specific way. There are two opposite directions (for
each saddle point) along which nearby trajectories approach the point at
an exponential rate and two other directions along which nearby
trajectories move away from it at an exponential rate.

We recall briefly some important notions from dynamical systems theory that
will be used in the present paper. Invariant manifold in a two-dimensional
flow is a material line, i.~e., it is composed of the same fluid particles
in course of time. To introduce the notion of stable and unstable manifolds
it is instructive to consider a steady flow around a hyperbolic point which
is the fixed point the fluid motion around which is unstable. So, a fluid
particle approaches the hyperbolic point along its stable (unstable)
invariant manifolds when $t \to + (-) \infty$. Even in simple time-periodic
flows, the stable and unstable manifolds may intersect each other
transversally creating so-called homoclinic and heteroclinic tangles where
the fluid motion is so complicated that it may be strictly called chaotic,
the phenomenon known as chaotic advection
\citep{Aref,Ottino,Samelson,Pierrehumbert,KP06,Wiggins}.
Close fluid particles in the tangles rapidly diverge providing
very effective mechanism for mixing. Stable and unstable manifolds
are important organizing structures in the flow because they attract and repel
fluid particles (not belonging to them) at an exponential rate and
partition the flow into regions with different types of motion. Thus, they are
transport barriers.

Stable and unstable manifolds are useful tools in studying realistic flows
modeling the ocean. In aperiodic flows it is possible to identify aperiodically
moving hyperbolic points with stable and unstable effective manifolds
\citep{Haller}. Unlike the manifolds in steady and periodic flows, defined
in the infinite time limit, the ``effective'' manifolds of aperiodic
hyperbolic trajectories have a finite lifetime. The point is that they
may play the same role in organizing oceanic flows as do invariant manifolds
in simpler flows. The effective manifolds in course of their life undergo
stretching and folding at progressively small scales and intersect each other
in the homoclinic points in the vicinity of which fluid particles move
chaotically. Trajectories of initially close fluid particles diverge rapidly
in these regions, and particles from other regions appear there. It is the
mechanism for effective transport and mixing of water masses in the ocean.
Moreover, stable and unstable effective manifolds constitute Lagrangian
transport barriers between different regions because they are material
invariant curves that cannot be crossed by purely advective processes.

Motion in any preselected region in a circulation basin may be considered as
a scattering problem in the sense that fluid particles come into the region
from outside and leave it sooner or later.
Passive particles are advected by an incoming flow into a mixing region,
where their motion may be chaotic, and then most of them are washed
away from that region. It is known in theory of chaotic scattering
that there exists an abstract chaotic invariant set in a bounded region
of phase space consisting of an infinite number of hyperbolic
particle trajectories that never leave the mixing region
\citep{Ott,Tel,BUP04,Budyansky}. If a particle belongs to the set at an
initial moment, then it remains in the mixing region forever (in theory).
Most of particles sooner or later leave the mixing region, but their
behavior is strongly influenced by the presence of the chaotic invariant set.
Each trajectory in the set and therefore, the whole set possesses stable
and unstable manifolds. Theoretically, these manifolds have infinite spatial
extent, and the tracer, belonging to the stable manifold, is advected by the
incoming flow into the mixing region and remains there forever.
The corresponding initial conditions make up a set of zero measure.
However, the particles that are initially close to those in the stable
manifold follow them for a long time, eventually deviate from them
and leave the mixing region along the unstable manifold. In Appendix A, we
illustrate these abstract mathematical concepts with the simple model of the
flow with a fixed point eddy embedded in a background steady flow with
the periodic tidal component \citep{BUP04,Budyansky} and explain how they
can be used to characterize inhomogeneous mixing in realistic eddy flows.

\section{Numerical Japan sea circulation model}

The Japan Sea is a deep marginal sea with shallow straits connected with
the East China Sea, Okhotsk Sea and North Pacific. The Japan Sea has
three deep basins, named as Tsushima and Yamato Basins in the southern
sea area and the deepest and largest Japan Basin in the northern
sea area. The paper is focused on simulation of the mesoscale dynamical
processes over the shelf and continental slope of the Japan Basin situated
in the northwestern area of the Japan Sea. The typical large scale
circulation over the Northwestern Japan Sea includes, two cyclonic gyres,
the cold Primorskoye Current streamed southwestward along the continental
slope of the Japan Basin and the warm northern current along the slope of
Japanese Islands. The sea domain in numerical experiments is characterized
by thin shelf along the continental coast of the Russian North Primorye
Region (Primorsky Krai), wide shelf of the Peter the Great Bay in the
South Primorye Region and steep continental slope in the whole sea area
adjacent to the northwest Japan Sea coast. The southwestern cyclonic gyre
over southern and central areas of the Japan Basin is simulated in the model
domain as a large scale circulation.

The MHI ocean circulation model developed by N.B.~Shapiro and E.N.~Mikhaylova
at the Marine Hydrophysical Institute, Sevastopol, Ukraine \citep{JSmodel}
is a set of 3D primitive equations under the hydrostatic and Boussinesq
approaches in Z coordinate system with a free surface boundary condition.
It belongs to a class of layered models, in which the sea consists of a
number of quasi-isopycnal layers. Interfacial surfaces between layers can
freely move up and down and layers can deform, physically vanish (outcrop)
and restore. The MHI model has been applied in
Refs.~\citep{Ponomarev,Trusenkova} for simulation of the Japan Sea large scale
circulation, as well as to simulate mesoscale dynamics in the northwestern
Japan Sea area adjacent to the Primorsky Krai coast. Equations of the MHI
model, vertically integrated within layers, are formulated at the beta-plane,
with the $x$-axis directed from the west to east and the $y$-axis directed
from the south to north. A specific feature of the MHI model is that
the density (buoyancy) of any layer is allowed to vary with space and time.

The present study is focused on simulation of mesoscale dynamics over the
continental slope and shelf in the closed sea area of the cyclonic gyre
occupying the southern and central area of the Japan Basin. The sea domain
is $39^{\circ}$N~-- $44^{\circ}$N, $129^{\circ}$E~-- $138^{\circ}$E
with the horizontal grid steps $2.5$~km along latitude and $2.5$~km
along longitude. The total number of the grid points is $210 \times 280$.
We set $9$ quasi-isopycnal layers including the upper mixed one. The bottom
topography is adopted from navigation maps. Islands can not be resolved
in coastline for the chosen mesh but they are represented in topography.
The area of the cyclonic gyre is suggested to be closed, and we use no slip
boundary conditions for current velocity at the sea domain contour
including sea coast, northern, eastern, and southern boundaries.

\begin{figure}[!htb]
\begin{center}
\includegraphics[width=0.48\textwidth,clip]{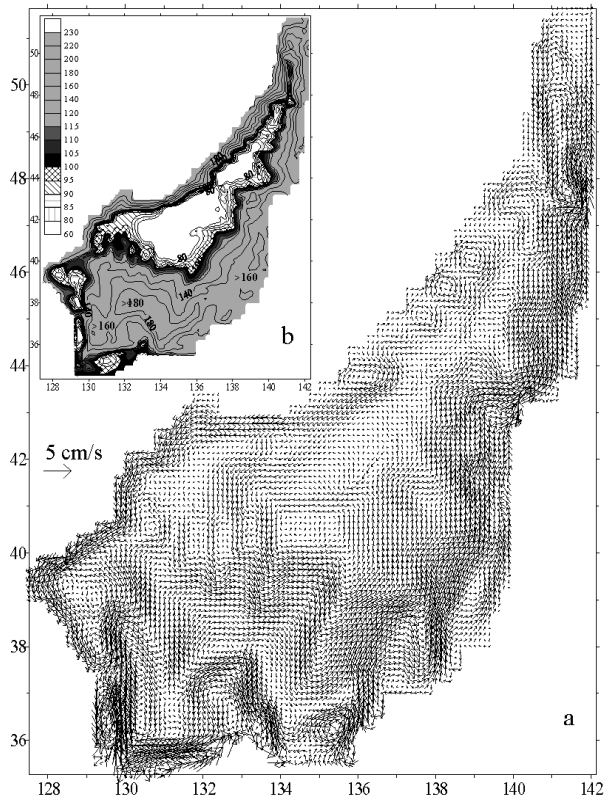}
\end{center}
\caption{(a) The velocity field in the whole basin of the Japan Sea in the
6th layer of the MHI model with the horizontal resolution $1/16^{\circ}$
and with initial temperature and salinity distributions estimated from the
CTD data of the oceanographic observations in summer 1999 \citep{Talley}.
(b) Depths of the main pycnocline (the interface between the 5th and 6th
layers of the model in August).}
\label{fig1}
\end{figure}

The initial conditions for realistic summer isopycnal interfaces, temperature
and salinity distribution in the model layers have been taken from oceanographic
survey 1999 \citep{Talley}. The MHI model has been integrated with the time step
of 4~min for a one year. The coefficients of quasi-isopycnal biharmonic
viscosity, harmonic viscosity, and diffusion used in the momentum and
heat/salt transfer equations have been varied correspondingly from
$10^{17}$~m${}^4$/s, $10^{7}$~m${}^2$/s and $0.4 \times 10^{7}$~m${}^2$/s
in the model spin up (60 days) to $10^{16}$~m${}^4$/s, $10^{6}$~m${}^2$/s
and $0.4 \times 10^{6}$~m${}^2$/s during other months of the warm period
of a year until mid November. During the winter convection, the coefficients
increase like in the spin up process. The quasi-isopycnal harmonic
viscosity is applied only near the domain boundary in a warm period
of a year and in the whole area in winter.

We simulate the nonlinear mesoscale eddy dynamics over the shelf, continental
slope, and Japan Basin taking into account realistic bottom topography and
daily mean external atmospheric forcing. The near-surface daily atmospheric
conditions have been set from the NCEP/NCAR Reanalysis. It includes short wave
radiation flux, wind stress, wind speed, air temperature and precipitation.

The numerical experiments with minimized coefficients of the horizontal
and vertical viscosity show the intensive mesoscale dynamics, particularly,
mesoscale variability of anticyclonic/cyclonic eddies and streamers over the
shelf and continental slope. The anticyclonic eddies, generated over the shelf
break and continental slope, move usually  southwestward along the slope
like the topographic Kelvin waves with prevailing phase velocity
of about $6$--$8$~cm/s. The spatial scale of the anticyclonic eddies
increases usually  near the Peter the Great Bay shelf where it exceeds
significantly the baroclinic Rossby deformation radius.

The current system and mesoscale dynamics over the continental slope and the
Peter the Great Bay shelf change substantially from summer to winter.
The strong northeastward boundary jet current is formed near the western
coast of the Peter the Great Bay from late October to November when the
monsoon is already changed from the summer type to the winter one.
We have simulated current velocity fields in the surface mixed layer
for the warm period with daily resolution in time.
This warm period is associated with strong baroclinic eddy activity
in the Northwestern Japan Sea.

Figure~\ref{fig1} demonstrates the velocity field in the whole basin of the
Japan Sea in the 6th layer of the model \citep{Ponomarev,Trusenkova} with
the horizontal resolution $1/16^{\circ}$ and with initial temperature and
salinity distributions estimated from the CTD data of the oceanographic
observations in summer 1999 \citep{Talley}. An annual run of atmospheric
conditions corresponds to the end of the 20th century. The mesoscale eddies
are generated due to the baroclinic instability in the main pycnocline
and manifest themselves in the upper layers as well. The velocity field in
Fig.~\ref{fig1} demonstrates clearly the mesoscale eddies along the coast
of the Prymorsky Krai region between $43^{\circ}$N and $46^{\circ}$N.
\begin{figure}[!htb]
\begin{center}
\includegraphics[width=0.45\textwidth,clip]{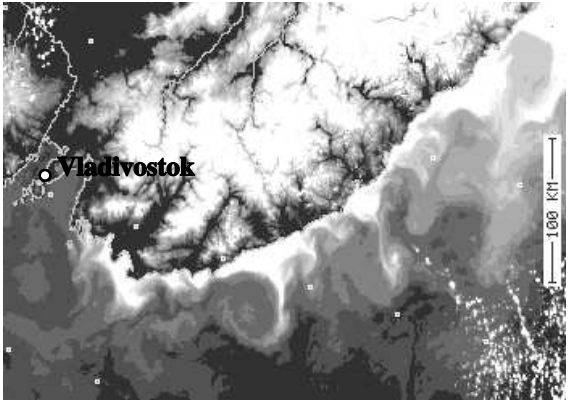}
\end{center}
\caption{Satellite image of the water surface temperature in the selected
region of the Japan Sea in the infrared range (NOAA AVHRR data, 15.~09.~1997).
Dark and white colors correspond to low and high temperatures, respectively.}
\label{fig2}
\end{figure}
\begin{figure}[!htb]
\begin{center}
\includegraphics[width=0.45\textwidth,clip]{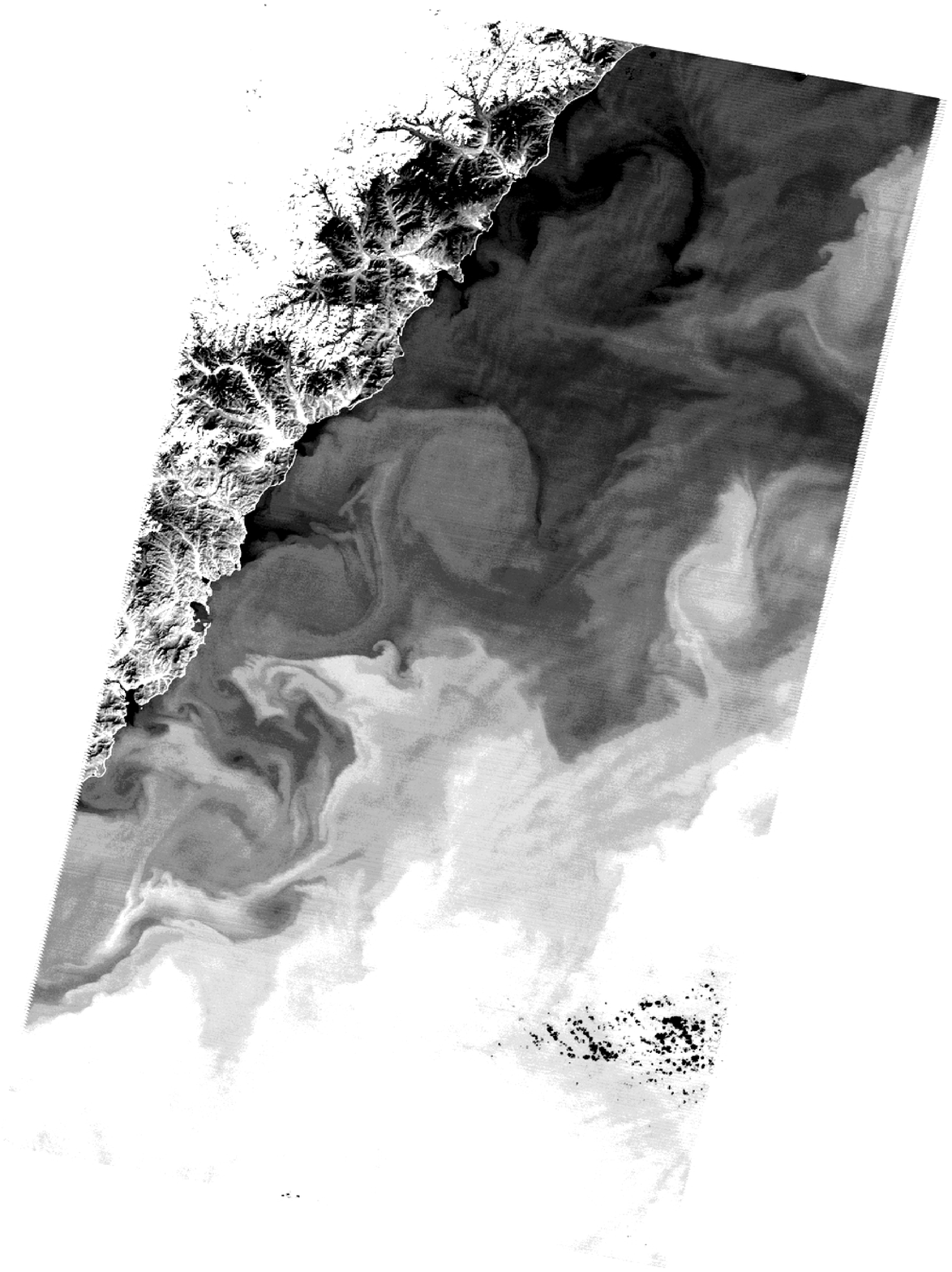}\\
\includegraphics[width=0.45\textwidth,clip]{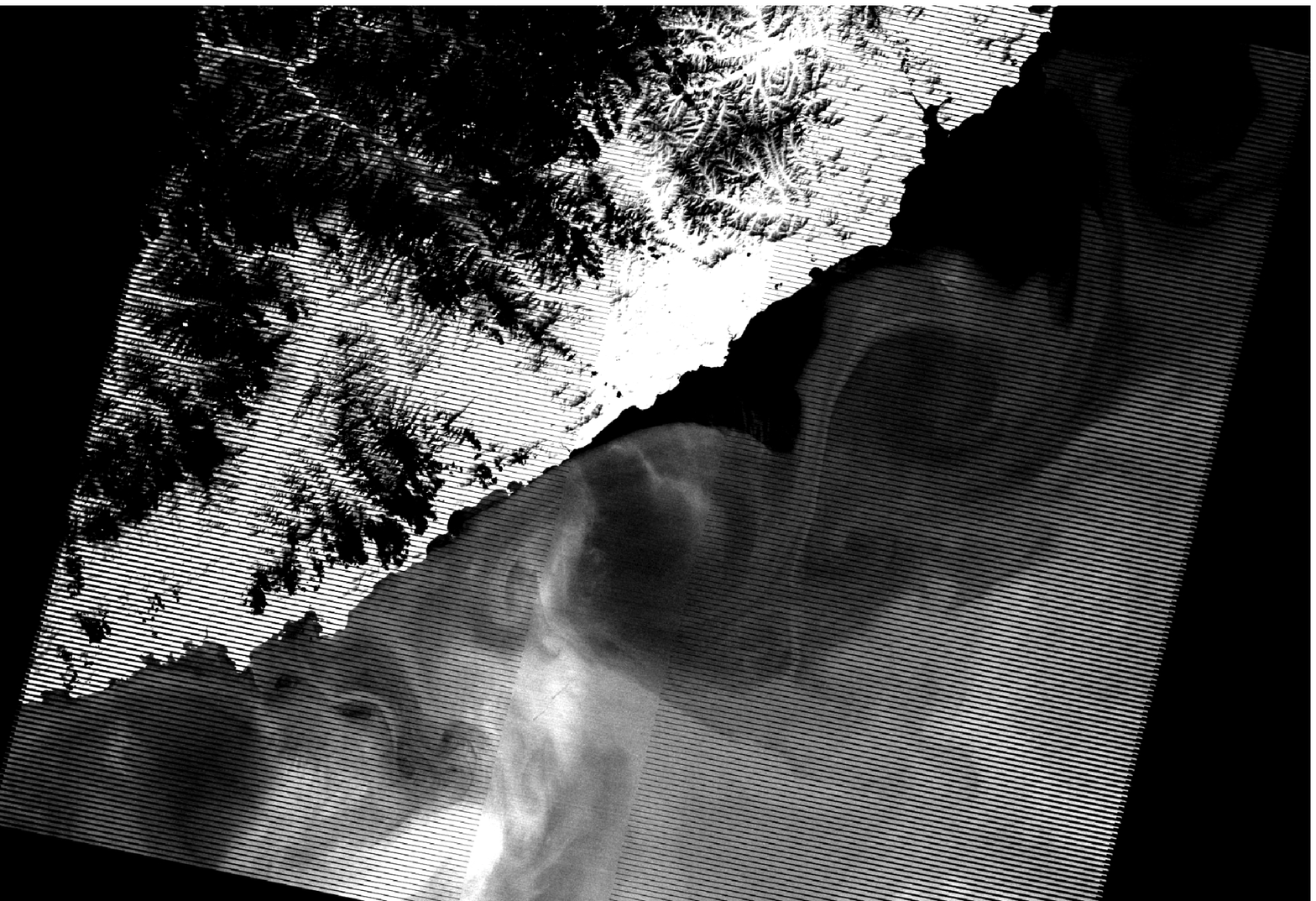}
\end{center}
\caption{The NOAA AVHRR  infrared images of the
vortex pairs in the mesoscale eddy street
associated with the Primorskoye Current. The upper panel:
Landsat-5~(TM) data (29.~09.~2007) with the resolution of
120~m. Coordinates: $43^{\circ}38^{\prime}$N-$45^{\circ}35^{\prime}$N,
$135^{\circ}10^{\prime}$E-$138^{\circ}14^{\prime}$E.
The lower panel: Landsat-7~(ETM+) data (14.~09.~2008)
with the resolution of 60~m.
Coordinates: $42^{\circ}10^{\prime}$N-$44^{\circ}10^{\prime}$N,
$133^{\circ}7^{\prime}$E-$136^{\circ}10^{\prime}$E.}
\label{fig3}
\end{figure}

\section{Lagrangian results}
\subsection{Finite-time Lyapunov exponents and repelling material lines}

In our analysis we focus on the region between latitudes
$41^{\circ}$N and $44^{\circ}$N
and between longitudes $130^{\circ}$E and $136^{\circ}$E comprising the
Primorskoye Current flowing to the southwest along the continental slope
of the Primorsky Krai (Russia). Satellite data demonstrate
anticyclonic mesoscale eddies of relatively small scale which are observed
in this area  directly over the steep continental slope.
The satellite image of the surface temperature in the infrared
range in the part of this region is shown in Fig.~\ref{fig2}.
Dark and white colors in the figure correspond to low and high temperatures, respectively.
The street of the anticyclonic eddies with the scale of 50~km is
visible along the coast of Primorsky Krai over the strip shelf and steep
continental slope of the Japan Basin. Centers of the eddies are situated
directly over the shelf break, 200~m depth. They
often form pairs of strongly interacting eddies visible at sea-surface temperature
satellite images. We demonstrate in Fig.~\ref{fig3} such vortex pairs
in the NOAA AVHRR Landsat-5~(TM) and Landsat-7~(ETM+) infrared images
in two different years with the resolution of 120 and 60~m, respectively
(http://glovis.usgs.gov/).
Each image shows the pair of two interconnected spirals
with a stagnation point between them.
The cores of the eddies are at the distance
$25 \div 30$~km from the coast of the Primorsky Krai.
The images also show the upwelling phenomena prevailing near the coast and
cold/warm streamers injecting the cold/warm water from the coastal
upwelling/offshore areas into the anticyclonic eddies.
We will study in this section such a vortex pair by
the Lagrangian methods using upper-layer current velocity fields from
numerical experiments with the
nonstationary hydrodynamic MHI sea circulation model.

\begin{figure*}[!htb]
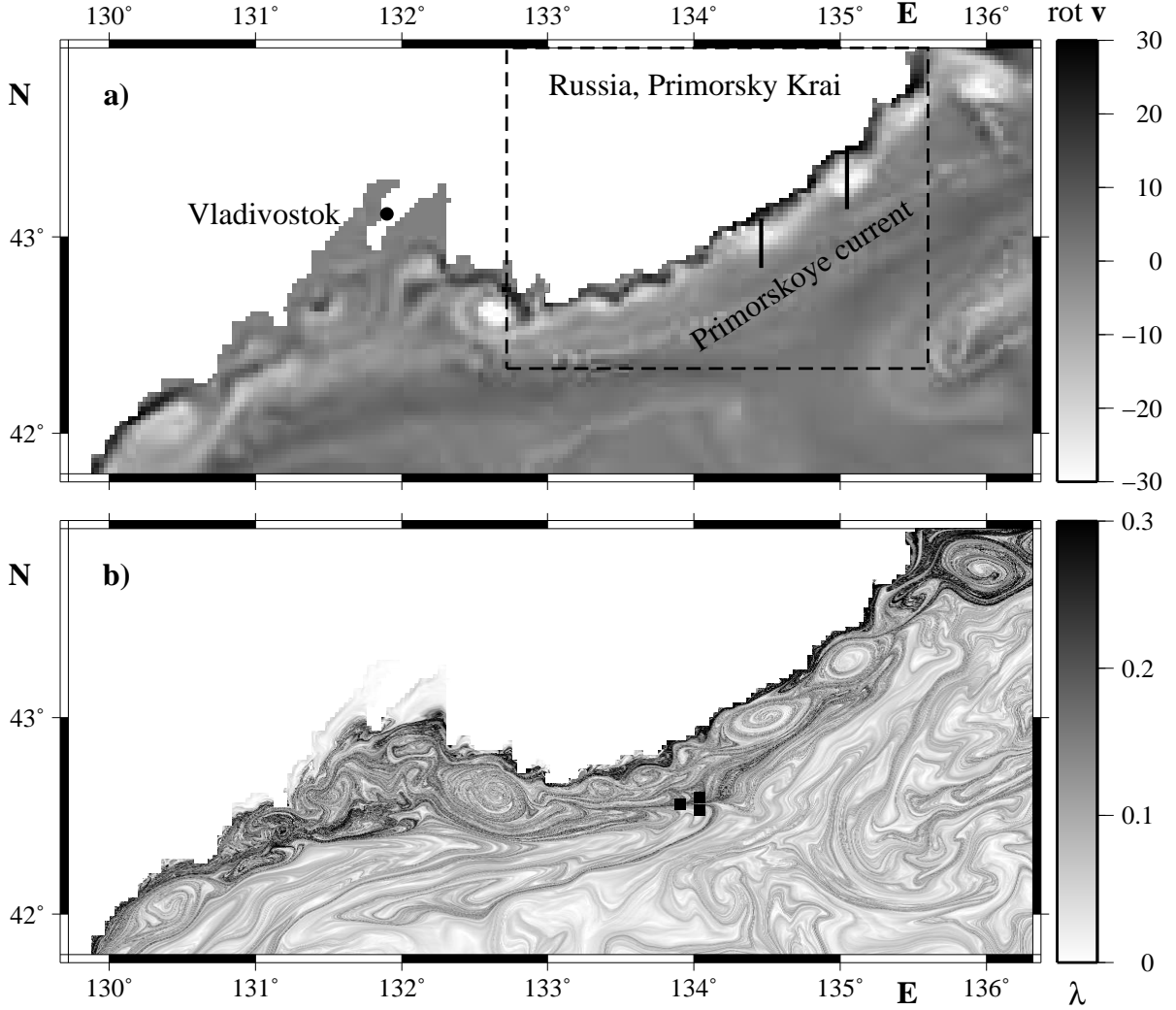

\begin{center}
\includegraphics[width=0.97\textwidth,clip]{fig4a.eps}\\
\includegraphics[width=0.97\textwidth,clip]{fig4b.eps}
\end{center}
\caption{(a) Snapshot of the vorticity field plotted vs initial particle's
positions. Color modulates the values of the vorticity
$\operatorname {rot} \mathbf{v}$. (b) Lyapunov synoptic map shows
the maximal FTLE, $\lambda$, vs initial particle's position. $\lambda$
is in units ${\rm days}^{-1}$. Integration time is 50 days.}
\label{fig4}
\end{figure*}

The mesoscale dynamics in the Primorskoye Current over the shelf and steep
continental slope, where the centers of the anticyclonic eddies
are situated directly over the strip shelf break, is associated with
the effect of
coastal Kelvin waves propagating southwestward along the shelf break.
The integration of both the MHI model and diagnostic model
\citep{Fyman}, using observed
temperature and salinity profiles from oceanographic
R/V surveys, shows the anticyclonic eddies with the scale of 50~km and
the centers over the shelf break moving downstream along the shelf and continental slope. The anticyclonic eddies
of the similar scale are simulated in the wide shelf of the Peter the Great Bay.
A snapshot of the vorticity field, $\operatorname {rot} \mathbf{v}$, on one of
the days in the first month of integration of the MHI model (Fig.~\ref{fig4}a)
demonstrates the complex pattern of mixing in that region with a number of anticyclonic eddies of
different sizes with negative vorticity.

In theory of dynamical systems, the Lyapunov exponents, $\lambda$'s, are known
to be quantitative criteria of chaotic motion in the asymptotic limit.
In practice, one computes Lyapunov exponents for a finite
time. The finite-time Lyapunov exponent (FTLE)
is the finite-time average of the maximal separation rate for a pair of
neighbouring advected particles. The FTLE at position $\mathbf r$
at time $\tau$ is given by
\begin{equation}
\lambda (\mathbf{r}(t))\equiv\frac{1}{\tau}\ln\sigma (G(t)),
\label{Lyapunov}
\end{equation}
where $\tau$ is an integration time, and $\sigma (G(t))$ denotes the
largest singular value of the evolution matrix $G(t)$ which governs
evolution of small displacements in linearized advection equations
(see Appendix B for the derivation of formula (\ref{Lyapunov})).

The FTLE is not an instantaneous separation rate, but rather measures
the integrated separation between trajectories. In real oceanic flows,
instantaneous streamlines can quickly diverge from
actual particle's trajectories. The FTLEs adequately describe
actual transport and mixing in the ocean because they are
derived directly from particle's trajectories.
They are especially useful in oceanography because they are mathematical
analogues of drifter launching in the ocean and characterize quantitatively
dispersion of water masses. Computing the FTLE field in a selected geographic
region, we get the map that contains information about mixing properties in
the region for a given period of time. Comparing the maps in different seasons,
we get an information about variability in the region.
Moreover, the Lyapunov maps enable to reveal
Lagrangian coherent structures hidden in the velocity field
including stable and unstable manifolds of finite-time hyperbolic
trajectories, large-scale transport barriers and eddies.
It is interesting that the FTLE is, in fact, an Eulerian quantity, but
in the same time it is a Lagrangian one because it is
derived from particle's trajectories.

A uniform grid of $1000 \times 1000$  particles is advected by
the numerically generated velocity field.
After $50$ days (starting on September, 15), the FTLEs are computed
using Eq.~(\ref{Lyapunov}). Spatial distribution of the FTLEs,
plotted against initial positions in Fig.~\ref{fig4}b,
may be called a Lyapunov  synoptic map. This map shows
that there is a large range of positive $\lambda$ values up to
$0.3$~days${}^{-1}$ which corresponds to Lyapunov mixing times
($e$-folding times) down to 3 days.
We would like to stress that the plot in
Fig.~\ref{fig4}b shows values of $\lambda$
against initial particle's positions accumulated for a rather
long time, 50 days.

The Lyapunov synoptic map in Fig.~\ref{fig4}b reveals a number of
structures. There are mesoscale anticyclonic eddies, forming the street along the
continental slope of the Primorsky Krai, which is easily visible on the Lyapunov map
and on the vorticity map (Fig.~\ref{fig4}a). The eddy's cores
are characterized by low values of the Lyapunov exponents. The particles
inside the cores tend to stay therein for a comparatively long time.
There are filaments that wind up around the eddy's centers in spirals which
reveal transport pathways of an ejection of water.

Moreover, there are very long
filaments, ridges of $\lambda$, corresponding to the largest Lyapunov exponents,
which are not associated  with any eddies.
The ridges with largest  values of $\lambda$,
sandwiched between the regions with smaller $\lambda$'s values, mean that
fluid particles placed initially on one of those ridges will experience in
the future strong hyperbolic behavior, i.~e., they will diverge from each
other at an exponential rate for the computed period of time. The ridges
reveal stretching directions of the velocity field. In the language of
dynamical systems theory, they approximate stable manifolds in a selected area.

\begin{figure}[!htb]
\begin{center}
\includegraphics[width=0.48\textwidth,clip]{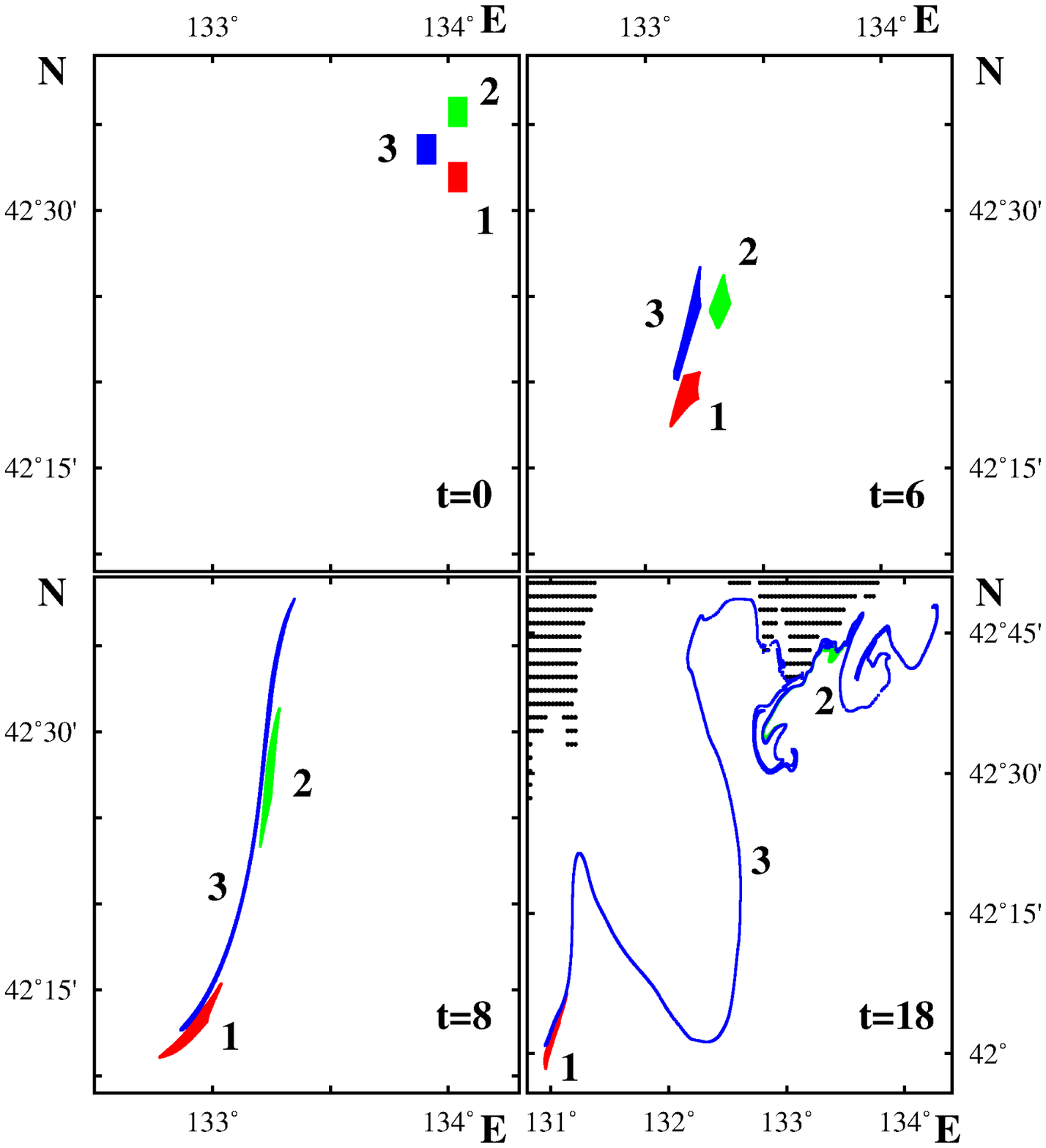}
\end{center}
\caption{Evolution of the coherent patches 1 (red) and 2 (green)
and the patch 3 (blue), which is strongly deformed, for 18 days.
Spatial scale in the panel at $t=18$ is different from the other ones. The size
of each patch is 6~km along the  latitude and 3~km along the longitude
with $250 \times 250$ particles in each one.}
\label{fig5}
\end{figure}
To demonstrate inhomogeneity of mixing in the region under consideration,
we compute the evolution of three fluid patches. The patch~3 was chosen
at the ridge of the Lyapunov map (it is the left rectangular among the three
ones marked in Fig.~\ref{fig4}b) with the centroid located initially at
($x_0=133^{\circ}54^{\prime}$E and $y_0=42^{\circ}34^{\prime}$N), whereas
the patches 1 ($x_0=134^{\circ}2^{\prime}$E and $y_0=42^{\circ}26^{\prime}$N)
and 2 ($x_0=134^{\circ}2^{\prime}$E and $y_0=42^{\circ}36^{\prime}$N) were chosen
nearby on the both sides of the ridge. Figure~\ref{fig5} compares their
evolution for 18 days. The patches~1 and 2 remain coherent for this
time but their fate is different: the patch~1 travels to the south-west,
whereas the patch~2 hits the coast to the north from
its initial location. Any ridge with largest  values of $\lambda$ serves,
in fact, as a transport barrier for waters on both its sides.

The behavior of the patch~3 in Fig.~\ref{fig5} is different.
It undergoes strong stretching and folding to be elongated for 18 days over
more than 600~km, almost two order of magnitude greater than the patches~1
and 2 do. This is because the patch~3 was initialized at the ridge of the
Lyapunov map which approximately corresponds to a stable manifold. As with any
patch placed near a stable manifold, the particles inside it align along
the associated  unstable manifold in course of time. It takes 5-6 days
for the patch~3 to reach the  unstable manifold. After that time, the
patch undergoes rapid stretching and folding.

\subsection{Lagrangian diagnostic tools for revealing eddy's structure}

In this section we focus on Lagrangian study of transport and mixing of passive
particles by the pair of strongly interacting anticyclonic
eddies in
the vortex street associated with the Primorskoye current. The satellite images of the
surface temperature (Fig.~\ref{fig3}) often demonstrate such pairs in
different years. The snapshots of the vorticity field
generated by the Japan Sea MHI model (Fig.~\ref{fig4}a) and the synoptic Lyapunov map
(Fig.~\ref{fig4}b) give the evidence of the vortex pairs in the same region. In addition to the
FTLE field, we propose new Lagrangian diagnostic tools for revealing eddy's
structure and eddy induced transport and mixing. To be concrete we consider
the most prominent pair of eddies in this region marked by the two straight lines
Fig.~\ref{fig4}a. They are also easily visible in the Lyapunov synoptic map
in Fig.~\ref{fig4}b.

\begin{figure*}[!htb]
\begin{center}
\includegraphics[width=0.97\textwidth,clip]{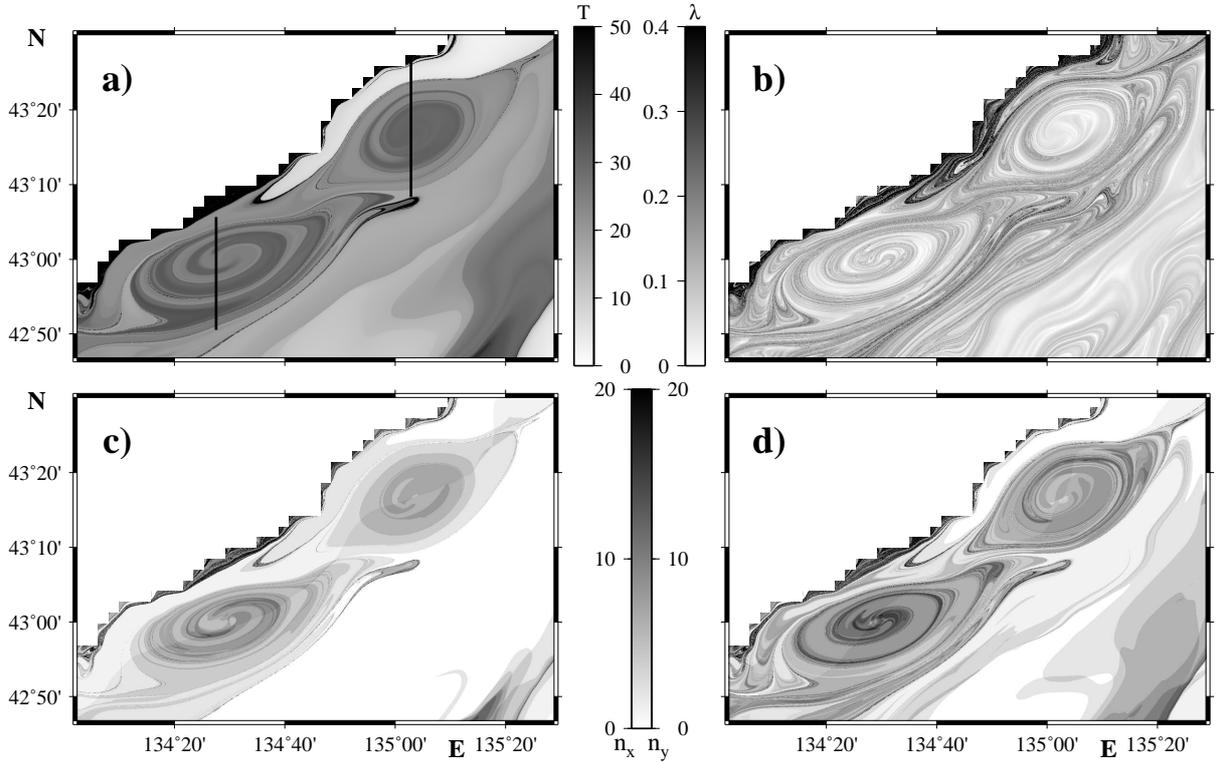}
\end{center}
\caption{(a) Exit-time map shows how long in days ($T$) it takes for a particle
with a given initial position to quit the
box shown in Fig.~\ref{fig4}a. (b) Lyapunov synoptic map of the region
shows the values of $\lambda$ vs initial particle's position. (c) and (d)
Maps of the number of times
advected particles change the sign of their
zonal and meridional velocities, $n_x$ and $n_y$, respectively,
unless they quit the box.}
\label{fig6}
\end{figure*}

It is worth to recall that the FTLE is an integrated quantity characterizing
the divergency
of nearby advected particles for a comparatively long period of time, 50 days
in our case. So, the FTLE synoptic map is a field of this quantity in geographic
coordinates which are initial positions of the passive particles.
Figure~\ref{fig6}b is a zoom of the Lyapunov synoptic map
in Fig.~\ref{fig4}b clearly demonstrating
the complex pattern of mixing of passive particles by the vortex pair selected.
The values of $\lambda$ form a spiral-like structure for each of
the eddies in the vortex pair. The very pair is surrounded by the ridges
of the largest values of the FTLE revealing stretching directions of the velocity
field which are stable manifolds of the hyperbolic trajectories of the
vortex pair. The spiral-like structure, in turn,
is the pair of spiral bands, one with largest values of $\lambda$ and the other
one with small FTLE values. The band with largest FTLE values is a collection of
initial particle positions which leave the corresponding eddy region in
course of time, whereas the band with minimal FTLE values marks those particles
that do not quit the eddy for 50 days. Thus, the shadowed spirals in
Fig.~\ref{fig6}b provide the Lagrangian information on the transport pathways
along which advected particles quit the corresponding eddy in course of time.
Using the velocity fields and the Lagrangian eddy patterns we can estimate
the general speed of the anticyclonic eddies moving southwestward along
the shelf break as equal to $2 \div 3$~cm/s, while the current velocity
in the upper mixed layer and on the sea surface is
about $10 \div 20$~cm/s. It corresponds to the mean speed of the
Primorskoye Current in the upper layer of $200 \div 400$~m
in the main pycnocline.

The FTLE map has a fine grained structure with a number of details.
To get more smoothed picture of mixing it is sometimes useful to
compute the so-called exit-time map. We distribute initially $10^6$ particles
in the box shown in Fig.~\ref{fig4}a and compute how long it takes for a
particle with given initial position to leave the box. The corresponding map,
which is more contrast than the Lyapunov one, is shown in Fig.~\ref{fig6}a.
It clearly demonstrates that
the northern eddy has a prominent central homogeneous spot (the eddy's core)
with large values of the exit times. The southern eddy is represented
by the spiral beginning in the eddy's core.

There is the quantity that may provide a representative picture of
eddy-induced advection, the number of times particles
wind around the eddy's center. It is difficult to compute that number
exactly for a large number of particles but it may be approximated by
the number of times a given particle changes the sign
of its zonal ($n_x$) and meridional ($n_y$) velocities unless it reaches
one of the borders
of the box shown in Fig.~\ref{fig4}a. We plot in Figs.~\ref{fig6}c and~d geographic
maps of those quantities where they are coded by the color. Both the maps
demonstrate clearly the spiral-like structure of the vortex pair.

\begin{figure}[!htb]
\begin{center}
\includegraphics[width=0.48\textwidth,clip]{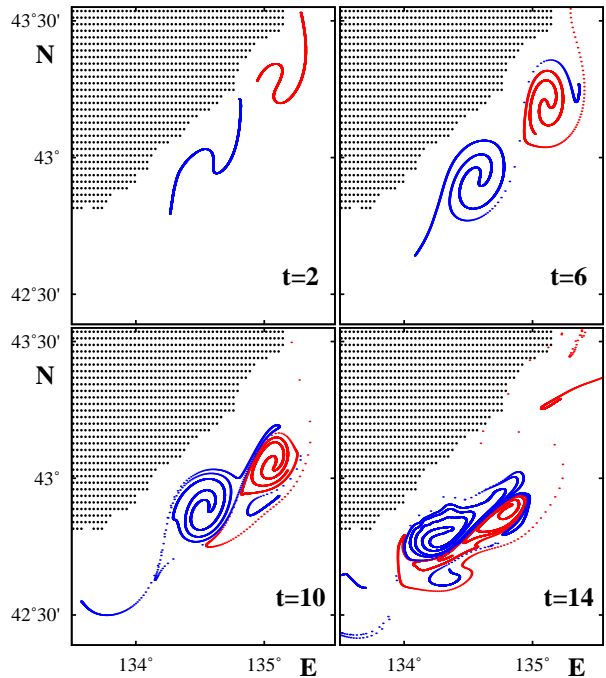}
\end{center}
\caption{Evolution of the two material lines (Fig.~\ref{fig4}a)
crossing the cores of the northern and southern eddies in the vortex pair.
The red (blue) color refers to the northern (southern) eddy.}
\label{fig7}
\end{figure}

To illustrate Lagrangian motion of particles in the vortex pair and
compare that motion in each of the eddies, we track the evolution of
material lines crossing the cores of the eddies. We take two straight
material lines in Fig.~\ref{fig4}a with $30000$ particles in each one
(the same lines are shown in Fig.~\ref{fig6}a) one of which
crosses the core of the northern eddy and the other --- the core of
the southern eddy.
Their evolution for 14 days is shown in
Fig.~\ref{fig7}. The dotted fragments in this figure appear
because of insufficiently large initial density of points.
The particles in the middle fragments of each line begin to rotate
anticyclonically
(the panel at $t=2$), form quickly the vortex pair (the panel at $t=6$)
and move downstream along with the eddy's cores winding around their centers.
The outer fragments of the lines elongate to the north and south
along the unstable manifolds of the eddy street (the panel at $t=10$).
At $t=14$, the northern eddy catches the southern one up and then both
the eddies move together.
\begin{figure}[!htb]
\begin{center}
\includegraphics[width=0.48\textwidth,clip]{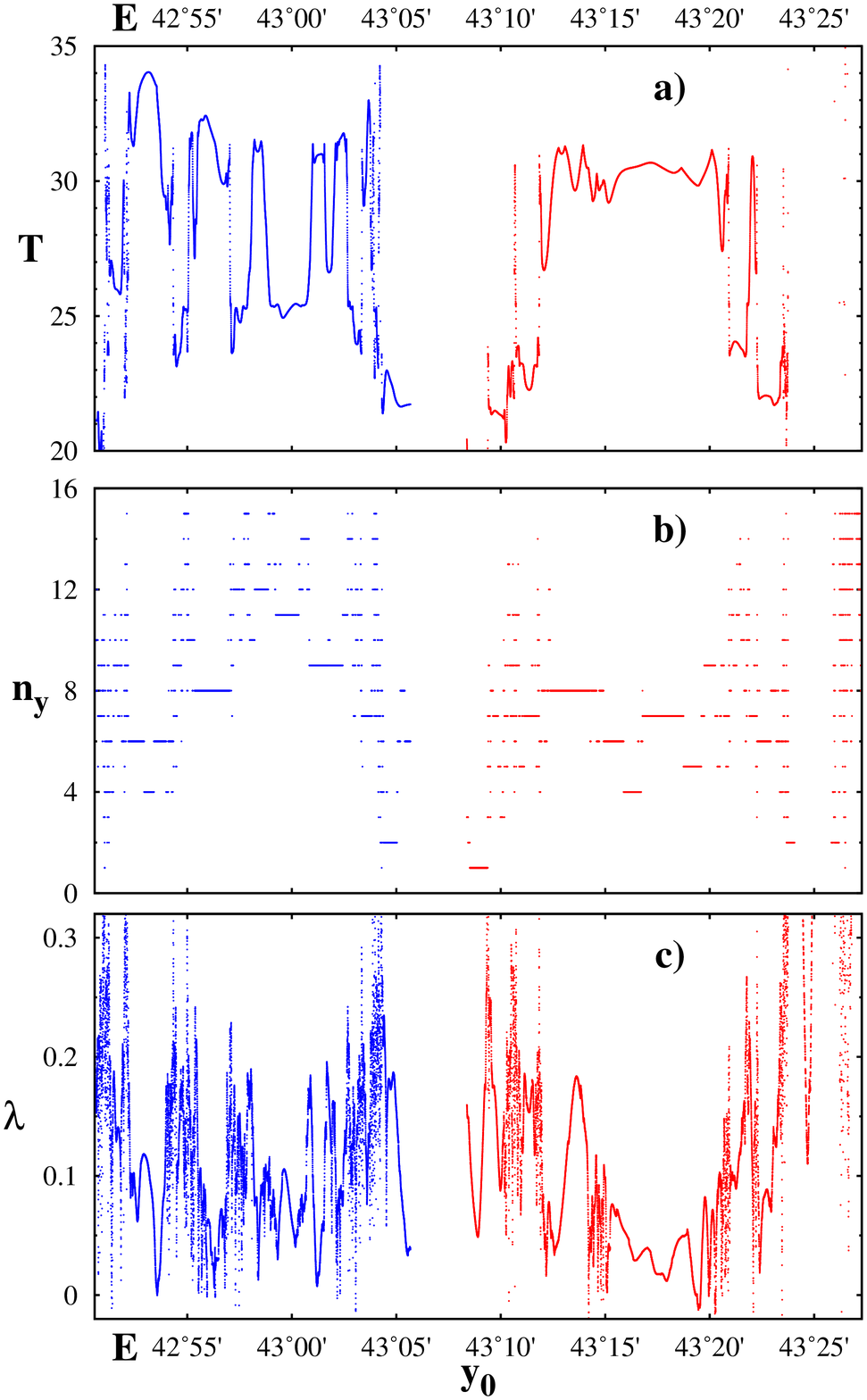}
\end{center}
\caption{ Evolution of the same two material lines as in Fig.~\ref{fig7}.
The left (blue) and right (red) columns
refer to the southern and northern eddies, respectively.
(a) Time of exit off the box $T$ in days vs initial
particle's latitude position $y_0$. (b) Number of times, $n_y$, the particles
change the sign of their meridional velocity before they
leave the box vs $y_0$.
(c) The corresponding maximal finite-time Lyapunov exponent vs $y_0$.}
\label{fig8}
\end{figure}

To give a detailed description of the structure of each eddy in the vortex pair we apply the method
of particle's scattering elaborated in Ref.~\citep{BUP04}.
This method is explained briefly in Appendix A of the present paper.
We take the same lines as shown in Figs.~\ref{fig4}a but compute now
dependencies of time of particle's exit off the selected box, $T$,
the number of times particles change the sign of their
meridional velocity, $n_y$, and the maximal FTLE on initial particle's
latitude $y_0$ (see Figs.~\ref{fig8}a, b and c, respectively).

The left column in Fig.~\ref{fig8}a with the latitudes
$y_0<43^{\circ}07^{\prime}$N
represents the data for the southern eddy and the right column with the latitudes
$y_0>43^{\circ}07^{\prime}$N --- the northern eddy.
Their comparison allows to see the difference between the two eddies in the vortex pair.
It is evident from Fig.~\ref{fig8}a that particles prefer to quit the southern eddy,
including its core, more or less periodically by portions. Each portion is
represented by a $\cup$-like segment of the $T(y_0)$ function
which consists of a large
number of particles with approximately the same time of exit and the
same number of changes of the sign of their meridional velocity
before leaving the selected box. It is seen in the plot
$n_y(y_0)$ in Fig.~\ref{fig8}b that the particles belonging to a given
$\cup$-like segment have the same values of $n_y$.
In difference from the southern eddy, particles quit the northern eddy's core practically
at the same time. In other words, the particles quit the southern eddy
by portions along spiral-like transport pathways, whereas the periphery of the
northern eddy exchanges water with the surrounding but its core
moves coherently as a whole for a long time.

Comparison of the left and right plots $n_y(y_0)$ in Fig.~\ref{fig8}b
gives an additional information on particle advection by the two eddies.
The particles in the core of the northern eddy change the sign of their
meridional velocity $4\div8$ times before leaving the selected box
(the right column), whereas the core of the southern eddy is
more inhomogeneous (the left column). It is confirmed by comparing
the plots $\lambda (y_0)$ for both the eddies in Fig.~\ref{fig8}c.

The scattering plots enable to identify in the most unambiguous manner
the eddy's core and its periphery. As an example let us consider the plots for
the northern  eddy shown in the right column in Fig.~\ref{fig8}.
The eddy's core is represented in the $T (y_0)$ plot in Fig.~\ref{fig8}a
by the smooth segment  of the length $\simeq 14$~km surrounded by the
inhomogeneous structures which should be attributed to the eddy's
periphery. The $\cup$-like segments of the function $T (y_0)$ beyond the
eddy's core represent the water masses evolving more or less coherently
and quitting the eddy by portions. This process is manifested in
Fig.~\ref{fig6} as spirals of the northern eddy which are less prominent
than those for the southern one. The motion of particles in the eddy's
periphery is erratic due to numerous intersections of effective
stable and unstable manifolds.

Figure~\ref{fig8}b gives the picture of circulation on submesoscales.
We show in Appendix A by means of the example with a simple vortex-current
system that both the trapping time for particles in the mixing
zone and the number of their full turns around the vortex have a
hierarchical fractal structure as functions of initial particle's position.
Both these functions (see Figs.~\ref{figB}a and b) are singular on
a Cantor set of initial conditions. Due to periodicity of that idealized
flow it became possible to explain in detail transport and mixing of
passive particles.

Numerically generated velocity fields in the ocean are, of course, aperiodic.
Moreover, we deal with moving eddies which are patches of nonzero vorticity.
So, we could not expect the simplified picture of scattering of material lines
by a moving eddy resembling that is shown in Appendix A. Nevertheless, we
find a kind of resemblance of the plots in Figs.~\ref{fig8}a and b with the
corresponding plots in Figs.~\ref{figB}a and b. The function $T(y_0)$ is a
graph with smooth $\cup$-like segments intermitted with badly resolved
ones. There is no, of course, singularities at the ends of smooth segments
because of lacking of periodicity of the velocity field and transient
character of effective stable and unstable manifolds. The plot $n_y(y_0)$
in Fig.~\ref{fig8}b is a hierarchy of epistrophes, but it is organized in
a much more complicated way as compared to the ones in Refs.~\citep{BUP04}
due to the same reasons. There is no evident fractal structure in both
the plots but there is a kind of self-similarity and strong difference
in behavior of neighbour particles whose exit times $T$ and the number
$n_y$ may differ by an order of magnitude for a rather short period of time.
Particles in each smooth segment in Fig.~\ref{fig8}a have approximately
the same number of changes of the velocity sign. So, each such segment
represents a coherent portion of the whole material line which is washed
away off the selected box simultaneously. The segments in Fig.~\ref{fig8}a, which
are not smooth, consist of a number of portions with their own fate and
very different values of $T$ and $n_y$.
\begin{figure}[!htb]
\begin{center}
\includegraphics[width=0.48\textwidth,clip]{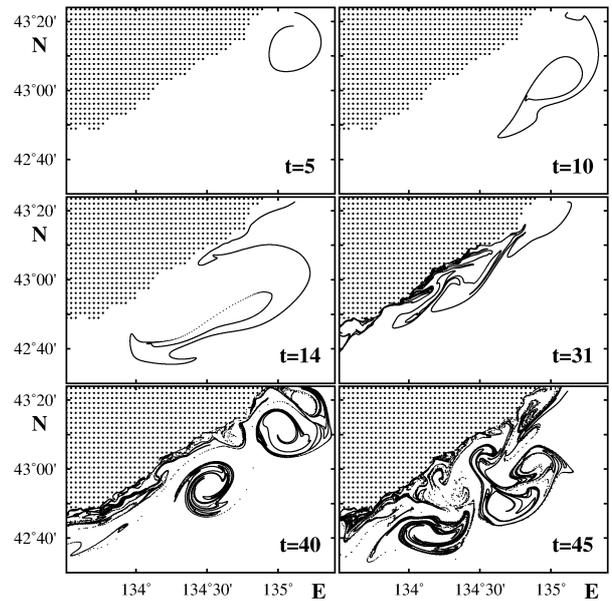}
\end{center}
\caption{Snapshots of the streakline obtained by injection of a dye for 45
days. The injection point ($x_0=135^{\circ}7^{\prime}$E,
$y_0=43^{\circ}23^{\prime}$N) was chosen at the periphery of one of the
eddies present there at the initialization time.}
\label{fig9}
\end{figure}

Important oceanographic information can be gained by computing so-called
streaklines in the active regions of the flow.
Streakline at the time moment $t$, passing through a point $(x,y)$, is a
curve composed of all the fluid particles which passed through that point
before the moment $t$. Injecting a dye into a point on the flow
plane, one can visualize the corresponding streakline. Recalling the
recent catastrophic oil spill at the bottom in the Gulf of Mexico, it
is evident that tracking the evolution of streaklines in numerically
generated or measured velocity fields
may provide useful information  about possible oil transport.

Figure~\ref{fig9} illustrates complex form of the streakline in the region under
consideration. The injection point ($x_0=135^{\circ}7^{\prime}$E,
$y_0=43^{\circ}23^{\prime}$N) was chosen to be at the periphery
of one of the anticyclonic eddies present there at the start of the injection.
That's why the injected particles begin to encircle the eddy anticyclonically
(see the panels on 5th and 10th days). The initial portion of the injected
particles moves downstream together with that eddy (the panel at $t=14$).
The dotted fragments of the streakline appear because of insufficiently
large initial density of points. The particles, released during the phase
without eddies near the injection point, begin to move downstream
(the panels at $t=10$ and $t=14$). Moving with the Primorskoye Current,
they successfully catch up the downstream eddies, and the streakline
begins to draw up all the anticyclonic eddies present in the region
(see the panels at $t=31$, $t=40$ and $t=45$). The resulting streakline
at $t=45$ gives an approximate image of the effective unstable
manifold of the whole region.

\section{Conclusion}

We have demonstrated in this paper that Lagrangian tools and methods
of dynamical systems theory can help to gain new information on surface
transport and mixing on both mesoscales and submesoscales in the ocean.
We have focused on the selected region of the Japan Sea comprising the coastal
Primorskoye Current with a street of anticyclonic mesoscale eddies.
Computing Lagrangian trajectories for a large number of particles advected
by the MHI numerical model, we have studied eddy-induced surface transport
and mixing in that region. We have developed the method to compute the FTLEs
for any velocity field and plotted the Lyapunov synoptic map with a high
resolution which can be used to quantify mixing processes.
The Landsat satellite infrared images, high resolution numerical experiments
with the MHI circulation model and Lagrangian modelling of the
mesoscale and submesoscale dynamics in the Primorskoye Current system
have shown strongly interacting mesoscale
anticyclonic eddies generated over
the shelf break and steep continental slope in the
northwestern area of the Japan Sea.

The main attention has been paid to Lagrangian study of transport and mixing
by a vortex pair of strongly interacting eddies which often occur in that
region in summer and autumn periods. We proposed new Lagrangian
diagnostic tools, the time of exit of particles off a selected box,
the number of changes of the sign of zonal and meridional velocities,
and computed synoptic maps for these quantities. Along with the Lyapunov
map, they have been shown to be able to reveal mesoscale eddies, meso- and
submesoscale filaments, repelling material lines, hyperbolic and
non-hyperbolic regions in the sea. In particular, we have found that the
eddies have a prominent spiral-like structure resembling the spiral
patterns at satellite images in that region.

Based on the theory of chaotic scattering, we developed the technique to
track evolution of clusters of particles, streaklines and material lines
and applied it to study in detail transport and mixing induced by the vortex
pair. The so-called scattering functions, dependencies
of the exit times and the number of times particles wind around the eddy's
center on initial particle positions, give us important oceanographic
information on eddy's structure and eddy-induced transport and mixing.
In particular, they allow to identify in the most unambiguous manner
the eddy's cores and periphery and to discover that the eddies release the
water to the surrounding by portions.

Lagrangian approach to transport and mixing problems seems to be perspective
because the results of computation of finite-time and finite-scale Lyapunov
exponents, exit times and winding numbers, tracking the evolution of
streaklines, material lines and patches of fluid particles will be more
and more realistic along with improvement of high-resolution numerical
models of the ocean circulation. The numerical results obtained within
that approach can be readily compared to the results of surface-drifter
and subsurface-float experiments which are rapidly becoming a common
experimental technique in oceanography
\citep{Garraffo,Molcard,Ozgokmen,Poje,Molcard06}.
This approach can be used as well for making predictions about possible
oil and other pollutant transport, biological productivity and other
applications. An interesting application of Lagrangian approach to study
marine ecosystem dynamics has been found recently in Ref.~\citep{PNAS}
where it has been demonstrated that frigatebirds may trace
precisely Lagrangian coherent structures in the Mozambique Channel
which are ridges in the field of the finite-size Lyapunov exponent.

\section*{Acknowledgments}

We would like to thank V. Dubina for providing us the satellite images
in Fig.~\ref{fig3} and anonymous reviewers for valuable comments.
The work was supported partially by the Program
``Fundamental Problems of  Nonlinear Dynamics'' of the Russian
Academy of Sciences, by the Russian Foundation
for Basic Research (project no. 09-05-98520) and by the
Prezidium of the Far-Eastern Branch of the RAS.

\appendix\section{Geometry of chaotic scattering of particles
in a simple vortex model}

It is instructive to demonstrate typical underlying structures
that govern complicated mixing and transport by ocean eddies
and introduce some special geometric concepts with an oversimplifed
model of the oceanic flow with a topographic eddy embedded in a
background steady flow with the periodic tidal component
\citep{BUP04,Budyansky}.

Passive particles, advected by a steady flow with the periodic component
directed along the $y$-axis, enter the mixing region where a fixed point
eddy with the singular point at $x=y=0$ is located and then wash out to an
outflow region. Let us put a large number of particles outside the mixing
region on a line segment, crossing the current, with the fixed value of
their $y_0$-coordinates and different values of $x_{0\,i}$. We compute
the time $T$, when the particles reach the line $y=6$, and the number of
particle's rotations $n$ around the point eddy before reaching this line.
It is a problem of chaotic scattering \citep{Ott}.
\begin{figure}[!htb]
\begin{center}
\includegraphics[width=0.48\textwidth,clip]{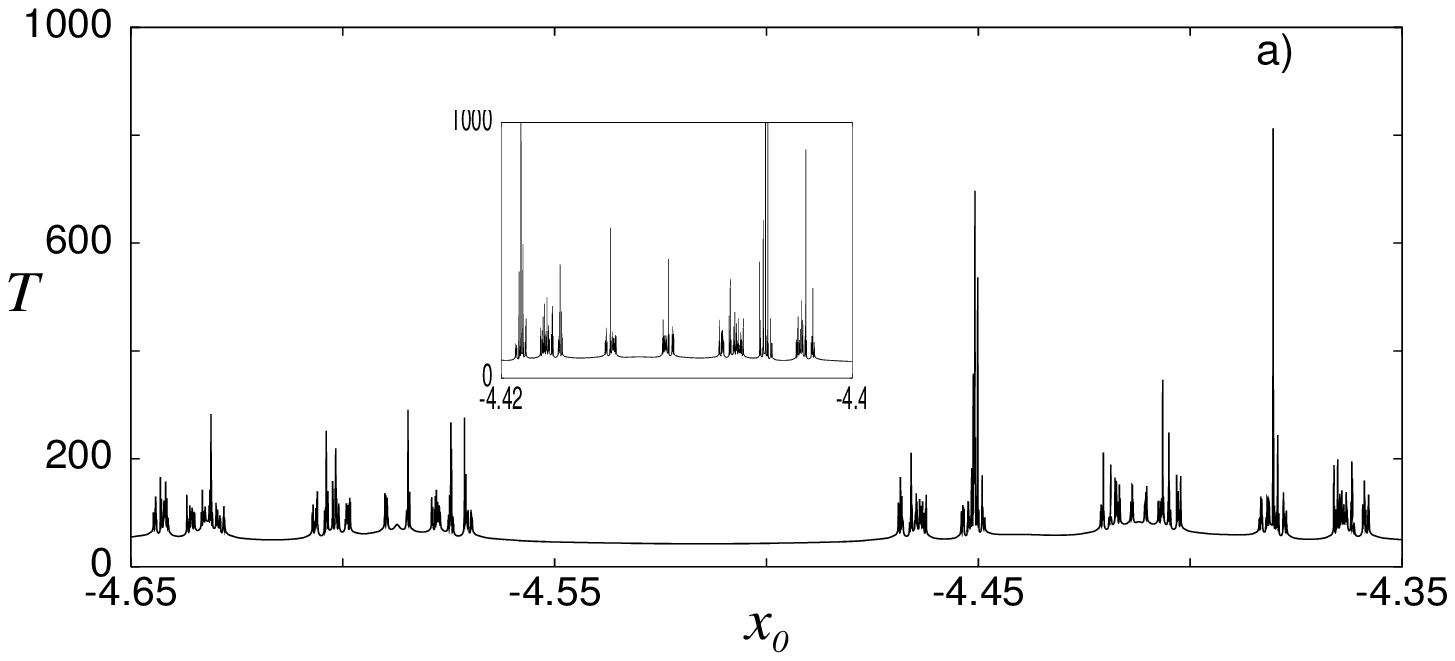}\\
\includegraphics[width=0.48\textwidth,clip]{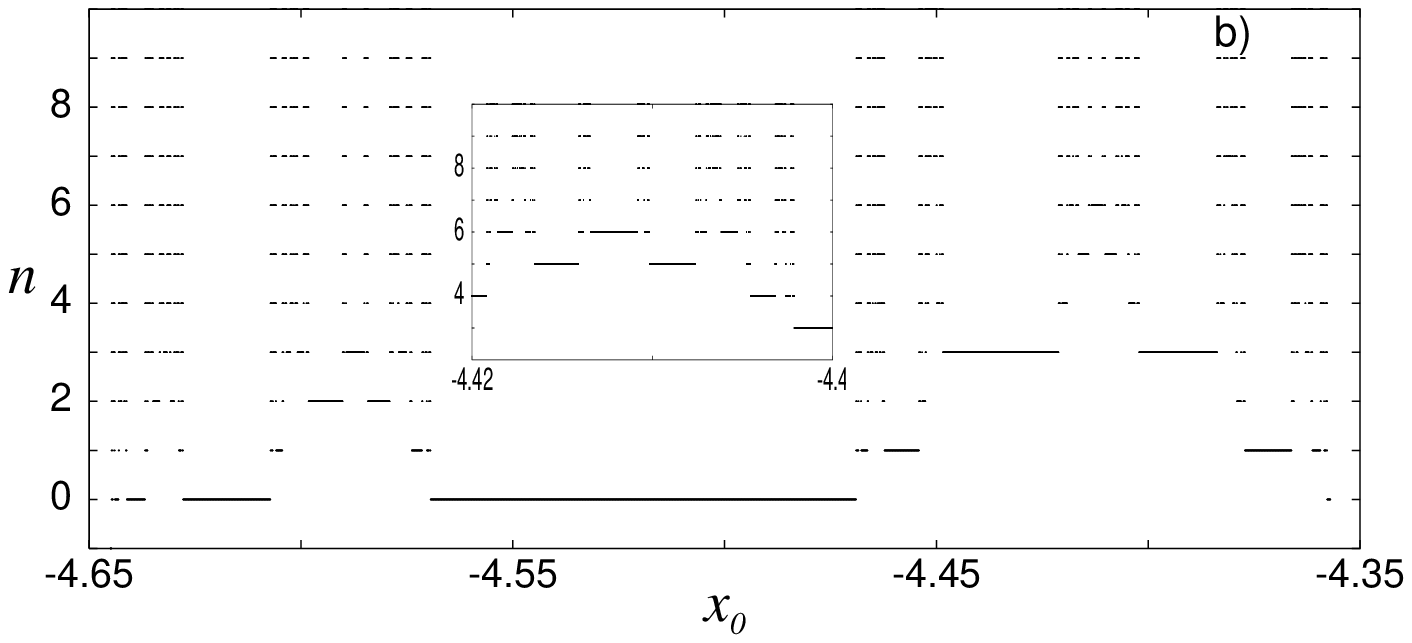}
\end{center}
\caption{(a) Fractal dependence of the trapping time $T$ on
initial particle's positions $x_0$ with the inset showing a 20-fold
magnification of one of the singularity zones. (b) Number of particle's
rotations $n$ around the point eddy before reaching a fixed line.
Mechanism of generating the fractal with magnification of a small
segment corresponding to the inset in the panel (a).}
\label{figB}
\end{figure}

Figure~\ref{figB} demonstrates the typical scattering
function $T(x_0)$ with an uncountable number of singularities which are
unresolved in principle. The inset in Fig.~\ref{figB}a shows a zoom by
the factor of 20 of one of these singularity zones. Successive magnifications
confirm a self-similarity of the function with increasing values of the
trapping times $T$. To give an insight into a mechanism of generating the
fractal, we plot in Fig.~\ref{figB}b segments of the initial string with a
large number of particles which are trapped in the mixing region after $n$
rotations around the fixed point eddy. After each rotation, a portion of
particles is washed out in the downstream region with $y\ge 6$. This process
resembles the mechanism of generating the famous Cantor set but is more
complicated. It is well known (see, for example, \citep{Ott}) that in
constructing the middle third Cantor set one takes the closed interval
$[0,\,1]$ and removes open middle third intervals from each of the intervals
remaining at each stage of the process. It is seen from Fig.~\ref{figB}b
that, starting with the given interval of initial points, different portions
of tracers are washed out by the flow from the intervals remaining after
each rotation around the vortex. Continuing in this way ad infinitum,
we get a Cantor set of remaining initial points.

Due to periodicity of that idealized flow it became possible to explain
in detail transport and mixing of passive particles \citep{BUP04}.
The point is that there exists an invariant chaotic set (saddle set)
$\Lambda$ \citep{Ott,Tel}. This set is defined as the set of all trajectories
(except for Kolmogorov--Arnold--Moser tori and cantori) that never leave
the mixing region. It consists of an infinite number of unstable periodic
and aperiodic (chaotic) trajectories. The particles belonging to $\Lambda$
remain on it forever. However, their measure is zero.

Each trajectory, belonging to $\Lambda$, and therefore the whole set has
stable, $\Lambda_s$, and unstable, $\Lambda_u$, manifolds which are infinite
material curves. The stable manifold of the chaotic set is defined as the
invariant set of trajectories approaching those in $\Lambda$ as
$t \to \infty$. The unstable manifold $\Lambda_u$ is defined as the stable
manifold corresponding to time-reversed dynamics. Following trajectories in
$\Lambda_s$, particles, advected by the incoming flow, enter the mixing
region and remain there forever. Particles that are initially close
to those in $\Lambda_s$ follow the corresponding trajectories, then deviate
from them and eventually leave the mixing region along the unstable manifold
$\Lambda_u$. So, if one chooses a material line in the incoming flow,
crossing the stable manifold $\Lambda_s$, it is expected that some
advected particles will stay in the mixing zone forever giving rise to
singularities in the $T(y_0)$ function. It means that this function has
a self-similar (fractal) structure with smooth $\cup$-like segments with
singularities at their ends intermitted with wildly oscillating fragments
with are unresolvable under magnification in principle.

Particles, which are chosen to be close to $\Lambda_s$ in the incoming flow,
enter along $\Lambda_s$ into the mixing zone where they remain for a long
time and eventually exit the region along the unstable manifold. The set
of segments with equal number of turns around the vortex, $n$, were called
``epistrophes'' in Ref.~\citep{BUP04} following to Ref.~\citep{Mitchel03}.
These epistrophes were shown to make up a hierarchy (Fig.~\ref{figB}b).
Each epistrophe converges to a limit point on the corresponding
material-line segment. The endpoints of each segment of the $n$th-level
epistrophe are the limit points of an $n+1$th-level epistrophe. The lengths
of segments in an epistrophe decreases in geometric  progression. The
common ratio of all the progressions $q$ is related to the maximal Lyapunov
exponent for the saddle points as follows: $\lambda=-\ln q/2\pi$.
The hierarchy of epistrophes in Fig.~\ref{figB}b determines transport of
particles, and its fractal properties are generated by the infinite sequences
of intersections of the advected material line with stable and unstable
manifolds of the chaotic invariant set $\Lambda$. The similar fractal-like
picture of chaotic transport and mixing has been found in more realistic
models for a periodically meandering jet current like the Gulf Stream and
the Kuroshio \citep{KP06,Chaos06,UlBP07}.

\section{Finite-time Lyapunov exponents}

In this Appendix we derive the formula (\ref{Lyapunov}) for the finite-time
Lyapunov exponents (FTLE) used in this paper to compute the Lyapunov synoptic
map for the Japan sea region selected (Fig.~\ref{fig4}b). The general problem
of particle's advection by a flow in an abstract $n$-dimensional space is
described by the $n$-dimensional system of nonlinear ordinary differential
equations in the vector form
\begin{equation}
\begin{gathered}
\mathbf{\dot x}=\mathbf{f}(\mathbf{x},t),\quad
\mathbf{x}=(x_1,\dotsc,x_n),\\
\mathbf{f}(\mathbf{x},t)=(f_1(x_1,\dotsc,x_n,t),\dots,f_n(x_1,\dotsc,x_n,t)).
\end{gathered}
\label{nonlinsys}
\end{equation}
The Lyapunov exponent at an arbitrary point $\mathbf{x_0}$ is given by
\begin{equation}
\Lambda(\mathbf{x_0})=\lim_{t\to\infty}\lim_{\Vert\delta \mathbf{x}(0)\Vert\to 0}\frac{\ln(\Vert\delta \mathbf{x}(t)\Vert/\Vert\delta \mathbf{x}(0)\Vert)}{t},
\label{lyap_def}
\end{equation}
where $\delta \mathbf{x}(t)=\mathbf{x_1}(t)-\mathbf{x_0}(t)$,
$\mathbf{x_0}(t)$ and $\mathbf{x_1}(t)$~ are solutions of the set
(\ref{nonlinsys}), $\mathbf{x_0}(0)=\mathbf{x_0}$. The limit exists,
is the same for almost all the choices of $\delta \mathbf{x}(0)$ and
has a clear geometrical sense: trajectories of two nearby particles
diverge in time exponentially (in average) with the exponent given by
the Lyapunov exponent.

Due to smallness of $\delta \mathbf{x}$ one can linearize the set
(\ref{nonlinsys}) in a vicinity of some trajectory $\mathbf{x_0}(t)$
and obtain the system of time-dependent linear equations~\citep{Greene}
\begin{equation}
\begin{pmatrix}
\delta\dot x_1\\
\hdotsfor{1}\\
\delta\dot x_n
\end{pmatrix}=J(t)
\begin{pmatrix}
\delta x_1\\
\hdotsfor{1}\\
\delta x_n
\end{pmatrix},
\label{linearization}
\end{equation}
where $J(t)$ is the Jacobian matrix of the system (\ref{nonlinsys}) along
the trajectory $\mathbf{x_0}(t)$
\begin{equation}
J(t)=
\begin{pmatrix}
\dfrac{\partial f_1(\mathbf{x_0}(t),t)}{\partial x_1}&\dots&\dfrac{\partial f_1(\mathbf{x_0}(t),t)}{\partial x_n}\\
\hdotsfor{3}\\
\dfrac{\partial f_n(\mathbf{x_0}(t),t)}{\partial x_1}&\dots&\dfrac{\partial f_n(\mathbf{x_0}(t),t)}{\partial x_n}
\end{pmatrix}.
\label{jacobian}
\end{equation}
Solution of the linear system (\ref{linearization}) can be found with the help of
the evolution matrix $G(t,t_0)$
\begin{equation}
\begin{pmatrix}
\delta x_1(t)\\
\hdotsfor{1}\\
\delta x_n(t)
\end{pmatrix}=G(t,t_0)
\begin{pmatrix}
\delta x_1(t_0)\\
\hdotsfor{1}\\
\delta x_n(t_0)
\end{pmatrix}.
\label{evol_mat}
\end{equation}
The evolution matrix obeys the differential equation which
can be obtained after substituting~(\ref{evol_mat})
into~(\ref{linearization})
\begin{equation}
\dot G=JG,
\label{evol_mat_diffur}
\end{equation}
with the initial condition $G(t_0,t_0)=I$, where $I$ is the unit matrix.
Any evolution matrix has the important property
\begin{equation}
G(t,t_0)=G(t,t_1)G(t_1,t_0).
\label{evol_mat_prop}
\end{equation}

One can write the singular-value decomposition  of the evolution matrix
as follows:
\begin{equation}
G(t,t_0)=U(t,t_0)D(t,t_0)V^T(t,t_0),
\label{evol_mat_svd}
\end{equation}
where $U$, $V$ are orthogonal and $D=\operatorname{diag}(\sigma_1,\dots,
\sigma_n)$ is diagonal.
The quantities $\sigma_1,\dots,\sigma_n$ are called singular values of
the matrix $G$. The Lyapunov exponents are defined via singular
values of the evolution matrix as follows:
\begin{equation}
\Lambda_i=\lim_{t\to\infty}\frac{\ln\sigma_i(t,t_0)}{t-t_0}.
\label{lyap_def_sigma}
\end{equation}
Quantities $$\lambda_i(t,t_0)=\frac{\ln\sigma_i(t,t_0)}{t-t_0}$$
are called finite-time Lyapunov exponents~\citep{Okushima}.
Thus, the FTLE is the ratio of the logarithm of the maximal possible
stretching in a given direction to a time interval $t-t_0$.

Let us consider now our specific problem of particle's advection on a plane
with $2\times 2$ evolution matrix and the singular-value decomposition
\begin{multline}
G=UDV^T \Rightarrow
\begin{pmatrix}
a&b\\c&d
\end{pmatrix}
=
\begin{pmatrix}
\cos\phi_2&-\sin\phi_2\\
\sin\phi_2&\cos\phi_2
\end{pmatrix}\times\\
\begin{pmatrix}
\sigma_1&0\\
0&\sigma_2
\end{pmatrix}
\begin{pmatrix}
\cos\phi_1&-\sin\phi_1\\
\sin\phi_1&\cos\phi_1
\end{pmatrix}.
\label{SVD2x2}
\end{multline}
Solution of these four algebraic equations has the form
\begin{equation}
\begin{gathered}
\sigma_1=\frac{\sqrt{(a+d)^2+(c-b)^2}+\sqrt{(a-d)^2+(b+c)^2}}{2},\\
\sigma_2=\frac{\sqrt{(a+d)^2+(c-b)^2}-\sqrt{(a-d)^2+(b+c)^2}}{2},\\
\phi_1=\frac{\operatorname{arctan2}{(c-b,\,a+d)}-\operatorname{arctan2}{(c+b,\,a-d)}}{2},\\
\phi_2=\frac{\operatorname{arctan2}{(c-b,\,a+d)}+\operatorname{arctan2}{(c+b,\,a-d})}{2},
\end{gathered}
\label{SVD2x2finsol}
\end{equation}
where function $\operatorname{arctan2}$ is defined as
\begin{equation}
\operatorname{arctan2}{(y,x)}=\left\{
\begin{aligned}
&\arctan{(y/x)}, &x\ge 0,\\
&\arctan{(y/x)}+\pi, &x<0.
\end{aligned}\right.
\label{arctg2}
\end{equation}

Equation (\ref{evol_mat_diffur}) can not be numerically integrated over
a large time because the elements of the corresponding evolution matrix
grow exponentially if one of the Lyapunov exponents is positive.
However, we can divide a large time interval on subintervals with the duration
which is less or order of the Lyapunov time, $t_\lambda=1/\lambda$,
and represent the whole evolution matrix as a product of evolution matrices
computed on these subintervals using the property~(\ref{evol_mat_prop}).
We compute this product and the corresponding singular values using the
GNU Multiple Precision Arithmetic Library~(http://gmplib.org) in order
to preserve the absolute precision of our representation of the
evolution matrix.

\bibliography{ourbib}{}
\bibliographystyle{model2-names}

\end{document}